\documentclass{article}

\usepackage{arxiv}

\usepackage[utf8]{inputenc} % allow utf-8 input
\usepackage[T1]{fontenc}    % use 8-bit T1 fonts
\usepackage{hyperref}       % hyperlinks
\usepackage{url}            % simple URL typesetting
\usepackage{booktabs}       % professional-quality tables
\usepackage{amsfonts}       % blackboard math symbols
\usepackage{nicefrac}       % compact symbols for 1/2, etc.
\usepackage{microtype}      % microtypography
\usepackage{lipsum}		% Can be removed after putting your text content
\usepackage{graphicx}
\usepackage{natbib}
\usepackage{framed,multirow}
\usepackage{doi}
\usepackage{amsmath}

\title{Multi-domain stain normalization for digital pathology: A cycle-consistent adversarial network for whole slide images}

\author{{\hspace{1mm}Martin J.~Hetz} \\
	Division of Digital Biomarkers for Oncology\\
	German Cancer Research Center (DKFZ)\\
	Heidelberg, Germany \\
	\texttt{martinjoachim.hetz@dkfz.de} \\
	\And
	{\hspace{1mm}Tabea-Clara ~Bucher} \\
	Division of Digital Biomarkers for Oncology\\
	German Cancer Research Center (DKFZ)\\
	Heidelberg, Germany \\
	\texttt{tabea.bucher@dkfz.de} \\
 	\And
	{\hspace{1mm}Titus J. ~Brinker}\thanks{Corresponding author} \\
	Division of Digital Biomarkers for Oncology\\
	German Cancer Research Center (DKFZ)\\
	Heidelberg, Germany \\
	\texttt{titus.brinker@dkfz.de} \\
}

\renewcommand{\headeright}{Hetz et al., 2022}
\renewcommand{\undertitle}{Preprint}
\renewcommand{\shorttitle}{Multi-domain stain normalization with MultiStain-CycleGAN}

\hypersetup{
pdftitle={Multi-domain stain normalization with MultiStain-CycleGAN},
pdfsubject={q-bio.NC, q-bio.QM},
pdfauthor={Martin J. Hetz, Tabea-Clara Bucher, Titus J. Brinker},
pdfkeywords={Stain normalization, Adversarial networks, Domain shift, Histopathology},
}

\begin{document}
\maketitle

\begin{abstract}
	The variation in histologic staining between different medical centers is one of the most profound challenges in the field of computer-aided diagnosis. The appearance disparity of pathological whole slide images causes algorithms to become less reliable, which in turn impedes the wide-spread applicability of downstream tasks like cancer diagnosis. Furthermore, different stainings lead to biases in the training which in case of domain shifts negatively affect the test performance. Therefore, in this paper we propose MultiStain-CycleGAN, a multi-domain approach to stain normalization based on CycleGAN. Our modifications to CycleGAN allow us to normalize images of different origins without retraining or using different models. We perform an extensive evaluation of our method using various metrics and compare it to commonly used methods that are multi-domain capable. First, we evaluate how well our method fools a domain classifier that tries to assign a medical center to an image.  Then, we test our normalization on the tumor classification performance of a downstream classifier. Furthermore, we evaluate the image quality of the normalized images using the Structural similarity index and the ability to reduce the domain shift using the Fréchet inception distance. We show that our method proves to be multi-domain capable, provides the highest image quality among the compared methods, and can most reliably fool the domain classifier while keeping the tumor classifier performance high. By reducing the domain influence, biases in the data can be removed on the one hand and the origin of the whole slide image can be disguised on the other, thus enhancing patient data privacy. 
\end{abstract}

% keywords can be removed

\section{Introduction}
\label{sec:introduction}
\begin{figure*}[!htb]
\centering
\includegraphics[width=.75\textwidth]{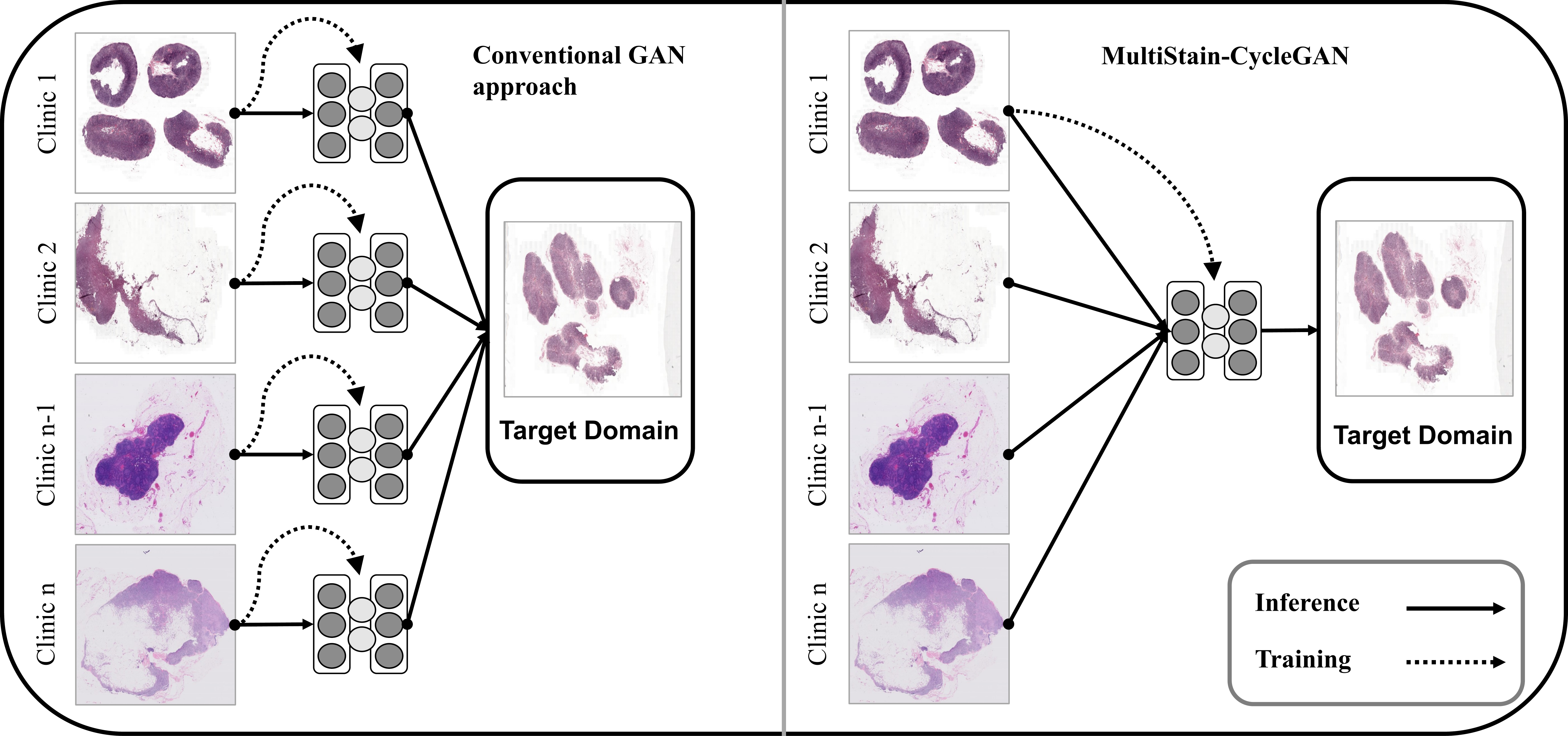}
\caption{
Left: Stain normalization based on conventional GAN approaches. For each staining a separate model has to be trained to normalize many stainings to the target domain. Right: Stain normalization with MultiStain-CycleGAN, which is trained on one staining and can normalize any H\&E staining of the same tissue type. Dotted arrows indicate the data needed for training the respective model, normal arrows show the inference path.  
}
\label{fig:gan_normalization}
\end{figure*}
The gold standard of cancer diagnosis is histopathologic investigation of tissue. This involves microscopic examination of dissected and stained tissue to examine signs and characteristics of specific diseases. During dissection, the tissue is fixed in formaldehyde and embedded in paraffin \citep{Chatterjee2014-sy}. The following staining of the tissue sections serves to highlight the various structures and cells in the tissue \citep{Ghaznavi2013-zs}. For light microscopy, the tissue sections are routinely stained with hematoxylin and eosin (H\&E) \citep{Alturkistani2015-wz}. The staining of the tissue helps the pathologist to make diagnoses based on certain features such as cell morphology or the arrangement of cells \citep{Gurcan2009-pd}. Staining mainly depends on the formulation of the stain and the application time among other pipeline-dependent aspects \citep{Kothari2014-fu}. 
After the tissue is stained, the slides are reviewed by pathologists. Both diagnosis and tumor grading are performed with the goal of providing prognosis and treatment recommendations \citep{Farahani2015-sk}.
With the advancement of technology and the rise of whole-slide imaging, i.e., the digitization of high-resolution microscopic images, digital pathology has rapidly gained importance \citep{Alturkistani2015-wz}. Digitizing the slides allows the use of automated systems for detection, classification or segmentation of a desired entity. Deep Learning (DL) and Convolutional Neural Networks (CNN) emerged as an effective tool in automatic image analysis, using big data to parameterize models that no longer rely on hand-crafted features \citep{LeCun2015-yu}. By using DL-based algorithms, pathologists can be relieved of monotonous and repetitive tasks by supporting them \citep{Echle2021-ek}. DL has already been successfully tested in various clinical tasks, such as tumor detection \citep{Cruz-Roa2017-ky}, mitosis detection \citep{Saha2018-cb,Albarqouni2016-ta}, grading of cancer \citep{Bulten2020-bz}, predicting the lymph sentinel status \citep{Kiehl2021-um}, patient survival estimation \citep{Wessels2022-ly} or tumor subtyping \citep{Wang2020-ai}. \\
The appearance of slides is highly variable as a result of different scanners, staining techniques, and laboratories \citep{Ciompi2017-qd}. This variability is not a problem for pathologists \citep{Bancroft2008-gk}, but current deep learning algorithms have significant issues with this \citep{Ciompi2017-qd}. Even very small changes in the image can lead to large deviations in performance for DL-based algorithms \citep{Kurakin2016-ti}. This is especially true for domain shifts, where a different input data distribution can lead to reduced performance and in turn potentially harm the patient \citep{Stacke2021-nu}. As early as 1994, Lyon et al. postulated that the standardization of dyes and stains will play an increasingly important role in the future \citep{Lyon1994-dy}. However, not all differences are due to non-standardized processes and thus not all variations are avoidable \citep{Niethammer2010-tt}. In conclusion, every tissue source site e.g. medical center or clinic, has a distinct signature due to biological variations in patients treated at various centers, specimen acquisition, staining, and digitization. On the one hand, this center-specific signature leads to biases in the data, which many algorithms suffer from; on the other hand, it can be used to determine the source of a whole slide image (WSI). The origin of the WSI can then be used to draw conclusions about the patient demographics, such as age, nationality and ethnicity. DL is able to determine the origin of a WSI with high accuracy \citep{Howard2021-lw}. This means that patient data privacy is no longer guaranteed and also enables the misuse of this information. To integrate DL into the work of pathologists and clinicians, methods that make these algorithms robust and stain-invariant must be developed, as well as new approaches that can remove the center-specific signature. Stain normalization is one possible method for achieving this objective.

\subsection{Stain normalization}
\label{stain_normalization}

%\begin{figure*}[!htb]
%\centering
%\includegraphics[width=.75\textwidth]{slide_overview.pdf}
%\caption{
%Example slides for the different domains from the CAMELYON17 dataset. The images a)-e) show examples of tissue sections of the different centers with their different stainings: a) Canisius-Wilhelmina  Hospital (CWZ); b) Rijnstate  Hospital (RST); c) University  Medical  Center  Utrecht (UMCU); d) Radboud  University  Medical  Center (RUMC); e) Laboratory  of  Pathology  East-Netherlands (LPON) 
%}
%\label{fig:slides}
%\end{figure*}
Methods for normalization of histological slides have already been shown to be effective for some applications \citep{Ciompi2017-qd}. These normalization methods transform images $x$ of a domain $\mathcal{X}$ to look like they originated from a target domain $\mathcal{Y}$ or to match the appearance of a template image $y$. Because of this, the field of stain normalization is a very active field, which tries to map images from a source domain to a target domain. In this regard, Salvi et al. divides the field into the three areas: Global color normalization, color normalization after stain separation, and Deep Learning based normalization \citep{Salvi2021-nh}. 
Methods based on global color normalization apply procedures that use the statistics of a template image and obtain a transformation from it, such as the global color transformation using Principal Component Analysis (PCA) by \citep{Reinhard2001-qk} or histogram specification proposed by \citep{Gurcan2009-pd}. In methods based on stain normalization after stain separation, the individual staining components, usually hematoxylin and eosin, are separated. This technique makes use of the property that the two stains can be linearly separated by a transformation into optical density space \citep{Roy2018-ed}. Each pixel can then be calculated by the product of a stain color appearance matrix, which is acquired by a template image,  and the stain density map \citep{Salvi2021-nh}. The estimation of the appearance matrix can be based on singular value decomposition as described by \citep{Macenko2009-yu}, prior information \citep{Ruifrok2001-ll}, non-negative matrix factorization \citep{Vahadane2016-kj} or spectral matching \citep{Tosta2019-af}. 
More recent approaches rely increasingly on neural networks. Generative Adversarial Networks (GAN) are used to normalize stains and show promising results \citep{Zanjani2018-xi, Bentaieb2018-ro}. Following the work of \citep{Gatys2016-ss}, stain normalization is treated as neural style transfer, where the goal is to give an input image the appearance of a learned distribution of images. GANs are deep generative models which consist of two networks: A generator which generates images and a discriminator which tries to separate the generated images from the actual images corresponding to the distribution of the training data. The training is a minimax game in which the two models compete with each other. The goal of the generator is to generate images from a noise vector \(z\) such that the distribution of the generated images \(P_{G}(z)\) corresponds as closely as possible to the distribution of the training data \(P_{data}(x)\). The loss function can be described as:
\begin{equation}
\begin{split}
\min_{G}\max_{D} V\left(G,D\right) = {} &\mathbb{E}_{x\sim p_{data}\left(x\right)} \left[\log{\left(D_{x}\right)}\right] + \\
                             & \mathbb{E}_{z\sim p_{z}\left(z\right)} \left[\log{\left(1-D\left(G\left(z\right)\right)\right)}\right]. 
\end{split}
\end{equation}
The generator tries to minimize the loss function, while the discriminator tries to maximize it. The models are trained until they reach an optimum \citep{Goodfellow2014-rt}. 
\citep{Isola2017-nl} extends this approach by passing an image instead of a noise vector z as input to the generator. This allows images to be transferred to another domain, which is essential for the normalization of histological stains. However, the disadvantage of this approach is that image pairs are required. In histopathology, it is rare for the same slide to be re-stained multiple times, so this approach is of limited use. Salehi et al. circumvent this drawback by converting images to gray-scale images, thus generating image pairs synthetically \citep{Salehi2020-sg}. Due to the missing image pair problem, more work has been directed towards cycle-consistent adversarial networks (CycleGANs), proposed by \citep{Zhu2017-pb}, which no longer require image pairs by using cycle-consistency. Approaches based on CycleGAN are particularly suitable for the field of image-to-image translation in histopathology. CycleGAN and its variants have already demonstrated in several studies that they are suitable for stain-to-stain translation or normalization tasks \citep{Shaban2019-ce,De_Bel2021-io,Runz2021-bg,Zhou2019-is}.
\subsection{Multi-domain stain normalization}
\label{multi_domain}
The majority of previous literature focuses on transfer between two stains, but not on a many-to-one approach, which offers more flexibility in real-world settings. One-to-one approaches require to train a new model whenever stains from a new domain have to be normalized, see Fig. \ref{fig:gan_normalization}. Furthermore, they depend on the local availability of the data to perform the normalisation. In privacy-preserving settings such as federated or swarm learning, this is not the case. For this reason, we introduce MultiStain-CycleGAN, an unsupervised multi-domain capable stain normalization method based on CycleGAN. The method presented here follows a comparable approach as described \citep{Tellez2019-fo} but with the additional reconstruction of the input image to ensure the integrity of the image structure through the cycle consistency condition. In doing so, we reformulate the stain normalization task into an image-to-image translation task, in which heavily augmented input images are subjected to gray-scale conversion and transformed into the desired stain. The network learns to reconstruct images from gray images with different contrasts, which have the appearance of the learned staining and thus perform stain normalization. Our method is tested for several properties including tumor classification, domain classification, image quality of generated images and distribution shift after normalization. 
Our contributions can be summarized to:
\begin{itemize}
  \item Developing MultiStain-CycleGAN, a robust deep learning  based multi-domain approach for stain normalization without the need for retraining to normalize untrained stainings. 
  \item Achieving low accuracy on tissue source site classification to remove spurious domain factors and thus improving data privacy  while generating images with the highest SSIM index and retaining tumor prediction accuracy
  \item Detailed analysis of the studied normalization methods in terms of their ability to improve a downstream task, their image quality and their ability to disguise the origin of the images, as well as the influence of different data augmentation intensities. 
\end{itemize}

The following parts of this work are structured as follows: Chapter \ref{sec:materials} introduces the datasets and their attributes. In addition, it gives a description of how the data is acquired, preprocessed and stratified. Chapter \ref{sec:methods} explains stain normalization with CycleGAN and how we derived a multi-domain approach from it. Furthermore, the metrics used to measure the distribution shift and the image quality of the normalized images as well as the used neural networks are explained. Chapter \ref{sec:experimental_setup} describes the experimental setup and chapter \ref{sec:results} lists the results in detail. Afterwards the results are discussed and placed into context in chapter \ref{sec:discussion}. Chapter \ref{sec:conclusion} provides a concluding summary and identifies research gaps for future work.

\section{Materials}
\label{sec:materials}
\begin{figure*}[!htb]
\centering
\includegraphics[width=.75\textwidth]{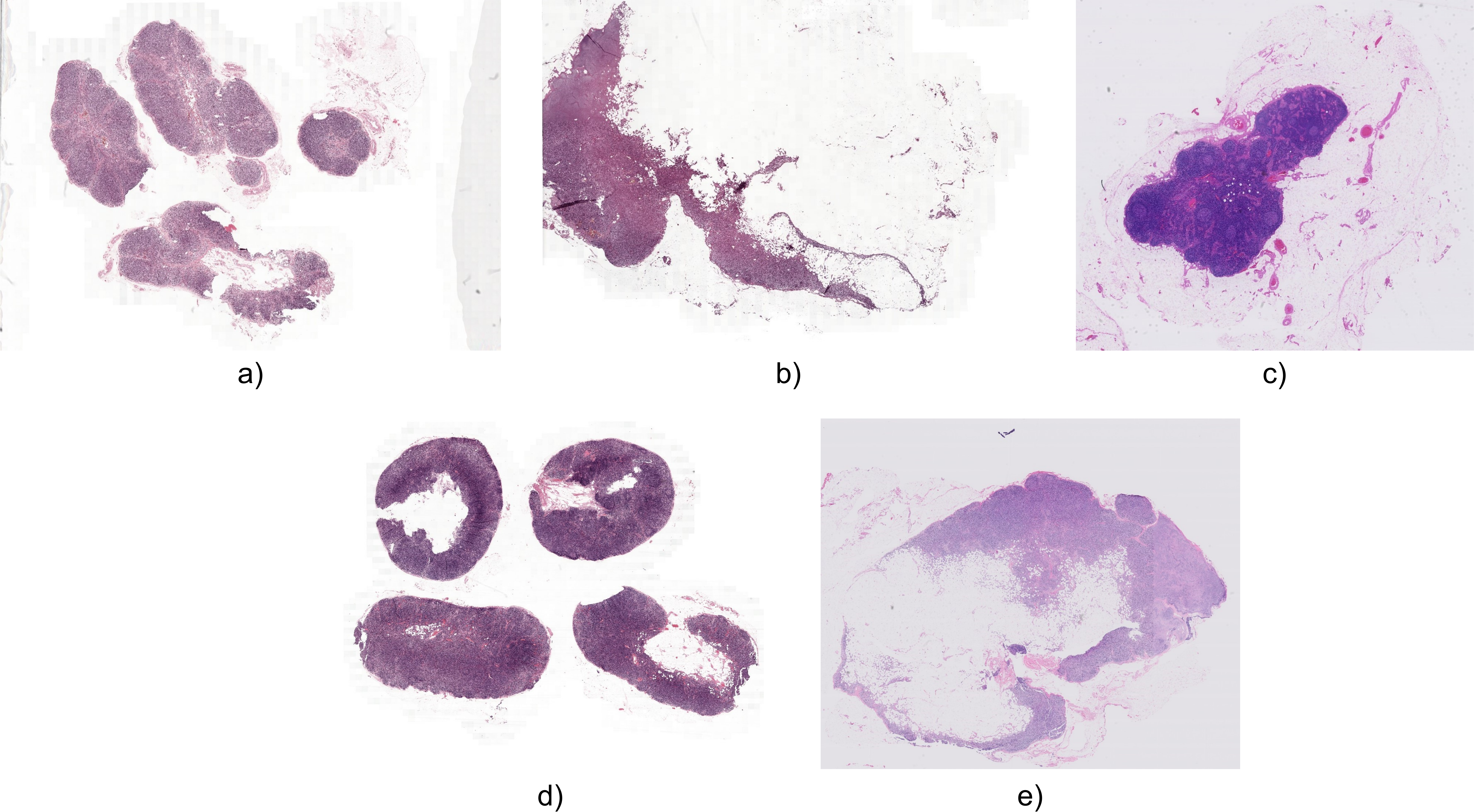}
\caption{
Example slides for the different domains from the CAMELYON17 dataset. The images a)-e) show examples of tissue sections of the different centers with their different stainings: a) CWZ; b) RST; c) UMCU; d) RUMC; e) LPON. 
}
\label{fig:slides}
\end{figure*}
The CAMELYON 17 Challenge dataset \citep{Bandi2019-rv} consists of whole slide images originating from five different medical centers, digitized with three different scanners. The five centers are: Radboud  University  Medical  Center (RUMC), Canisius-Wilhelmina  Hospital (CWZ), University  Medical  Center  Utrecht (UMCU), Rijnstate  Hospital  in  Arnhem  (RST) and Laboratory  of  Pathology  East-Netherlands (LPON). The dataset provides the domain shift we require for our investigation of stain normalization. The dataset encompasses a total of 500 breast tissue WSIs, 50 of them with annotations at lesion level. Examples for the appearance of the different stainings of the different domains are visualized in Fig. \ref{fig:slides}. All in all, this dataset allows us to make valid statements about the proposed normalization method and the quality of the normalized images. The 50 lesion-level annotated slides were primarily used to evaluate tumor classification, and the remaining slides were then used to classify the tissue submitting site, image quality, and determine the distance between the target and source domains. For the preprocessing of the tile-based approach, we utilized our self-developed publicly available pipeline\footnote{https://github.com/DBO-DKFZ/wsi\_preprocessing} for the extraction of tiles from WSIs and a subsequent filtering by tissue presence according to \citep{Khened2021-ti}.  We employed a configuration such that each tile had a spatial extent of \(64\times64\ \mu m\) and a resolution of \(256\times256\) pixels. Furthermore, we sampled the tiles with an overlap of 25\% for non-annotated and 50\% for annotated tissue. For the evaluation of our multi-domain stain normalization, we chose CWZ as the target domain. 
\subsection{Normalization Network}
Our MultiStain-CycleGAN required tiles from two of the five domains for training. In our experiments, we arbitrarily chose the two centers CWZ and LPON. For the training we used about 240 000 tiles from center CWZ and about 270 000 tiles from the center LPON. The tiles were extracted only from the slides containing lesion level annotations, i.e. 10 slides per center. 
\subsection{Tumor classifier}
For the tumor classifier, we used a 5-fold cross-validation. The folds were stratified according to the class 'tumor' or 'non-tumor'. In training, we only used data from the target domain with lesion-level annotations, with each fold containing 39 000 tiles, resulting in 195 000 for the total number of train images. Of these, a total of 24 000 tiles belong to the 'tumor' class. For the test set, the target domain CWZ was omitted because in this work we only focused on images with domain shift for tumor classifier performance. For the remaining centers, we decided to use 5000 stratified randomly drawn tiles of both classes, resulting in a total test set of 40 000 tiles for the tumor classifier.  
\subsection{Domain classifier}
Also for the training of the domain classifier we decided to use a 5-fold cross-validation, stratified by centers. Each training fold contains 290 000 tiles, in total about 1430 000 tiles, which were extracted from the 50 slides with lesion-level annotations. The test set consists of a total of 25 000 tiles, composed of 5 000 tiles per center. The test set is based on tiles from the 90 slides without lesion-level annotations, which have not been used in the evaluation before. 

\section{Methods}
\label{sec:methods}
This section introduces the general methods needed for this work such as CycleGAN and stain normalization. Then, we present how we derived MultiStain-CycleGAN and how we perform stain normalization. Furthermore, we explain how we conduct the study. Last but not least, we describe the used performance metrics.
\subsection{Cycle-consistent adversarial networks}
\begin{figure}[!t]
\centering
\includegraphics[scale=.135]{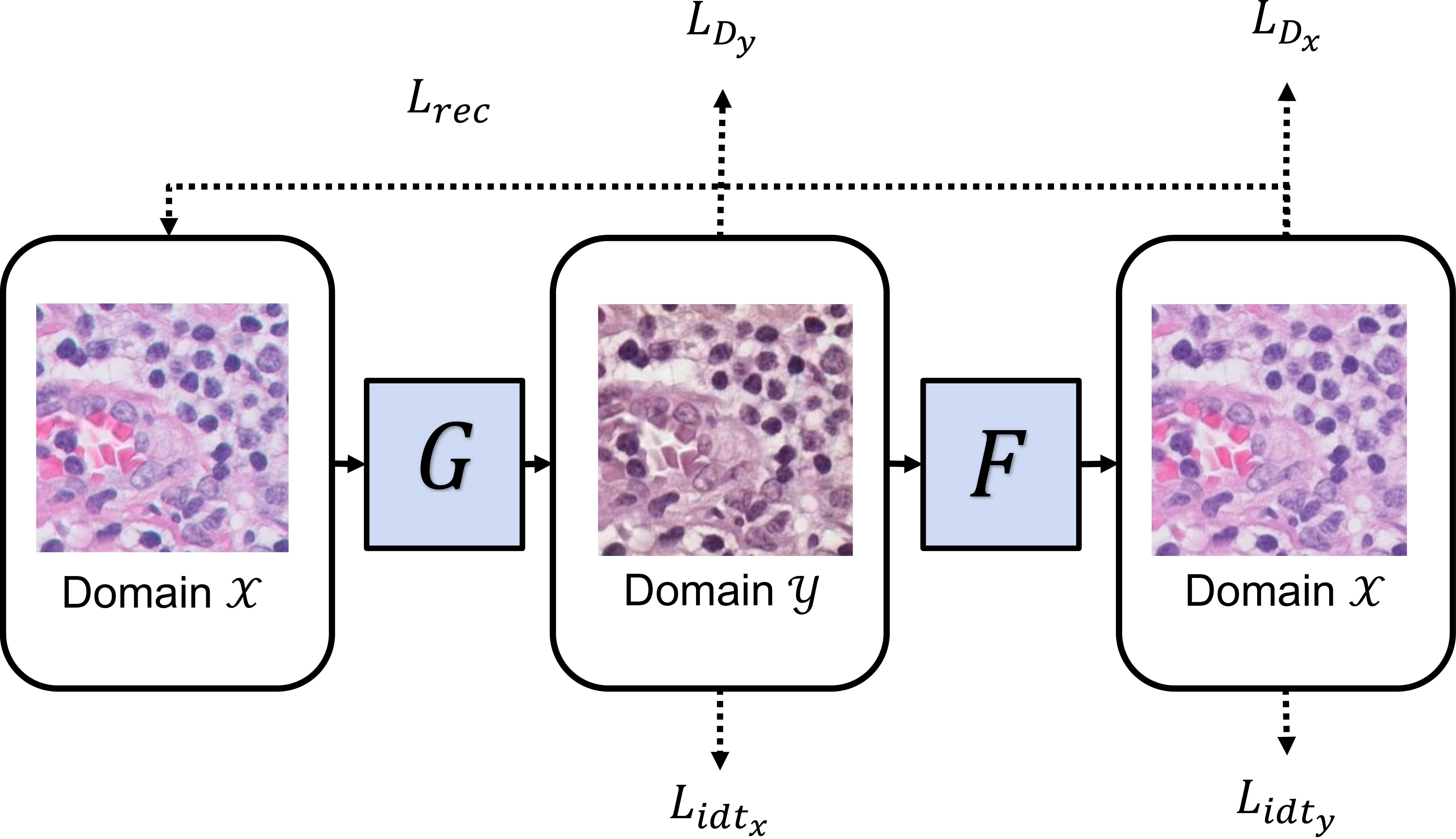}
\caption{The principle of image-to-image translation with CycleGAN proposed by Zhu et al. An image from a domain $\mathcal{X}$ is mapped to a domain $\mathcal{Y}$ by a generative model. After the mapping the image will be reconstructed into its original domain and the cycle-consistency loss is computed, enabling unpaired image-to-image translation. 
}
\label{fig:cycleGAN}
\end{figure}
Cycle-consistent adversarial networks proposed by Zhu et al. learn a mapping from a domain $\mathcal{X}$ to a domain $\mathcal{Y}$ using training data \(x_i \in X\) and \(y_i \in Y\). Where $X$ and $Y$ denotes the respective datasets of the domains. In Fig. \ref{fig:cycleGAN}, the two mappings are illustrated with \(G : \mathcal{X} \to \mathcal{Y}\) and \(F : \mathcal{Y} \to \mathcal{X}\). The domain-dependent adversarial discriminators \(D_x\) and \(D_y\) learn whether the input image is a generated image \(G(x)\) or \(F(y)\) or a sample \(x\) or \(y\) from the distribution of the training data \citep{Zhu2017-pb}. Here, the objective function contains several loss terms. Among them are adversarial losses \(L_{D_x}\) and  \(L_{D_y}\), to learn matching the distribution of generated and train data \citep{Goodfellow2014-rt} and a cycle-consistency loss \(L_{rec}\), which helps to preserve the structure of input images, as well as identity losses \(L_{idt_x}\) and \(L_{idt_y}\), which help to keep the color palette close to the input image \citep{Zhu2017-pb}.
The two adversarial losses are applied to the two mapping functions $G$ and $F$. For example, the adversarial loss for the function \(G : \mathcal{X} \to \mathcal{Y}\) with its discriminator \(D_y\) can be expressed as follows:
\begin{align}
\begin{split}\label{eq:gan}
L_{GAN}\left(G,D_y,X,Y\right) ={}& \mathbb{E}_{y\sim p_{data}(y)} \left[\log\left(D_{y}(y)\right)\right] + \\ &\mathbb{E}_{x\sim p_{data}(x)} \left[\log\left(1-D_y\left(G\left(x\right)\right)\right)\right].
\end{split}
\end{align}
Here, the generator $G$ tries to generate images $G(x)$ that appear as if they originate from the distribution $Y$, while the discriminator $D_y$ tries to distinguish the generated samples $G(x)$ from the real samples $y$. This process is repeated for the function \(F: \mathcal{Y} \to \mathcal{X}\) with \(L_{GAN}(F,D_x,Y,X)\)\citep{Zhu2017-pb}. 
Adversarial training can theoretically learn mappings $G$ and $F$ such that the generated images correspond to the distribution of the respective target domains $\mathcal{X}$ and $\mathcal{Y}$. Furthermore, with a sufficiently large capacity of the model, it is possible to map the same input to several different outputs in the target domain. To reduce the solution space, Zhu et al. suggest that the mapping functions should be cycle-consistent. This behavior is enforced by the cycle consistency loss with:
\begin{align}
\begin{split}\label{eq:cycle_consistency}
    L_{cyc} \left( G,F,X,Y \right) ={} & \mathbb{E}_{x\sim p_{data} \left( x \right)} \left[ \| F \left( G \left( x \right) \right) - x \|_1 \right] +\\
   & \mathbb{E}_{y\sim p_{data}\left(y\right)} \left[ \|G \left(F\left(y\right)\right) - y \|_1 \right].
\end{split}
\end{align}
For certain tasks, including stain normalization, it is useful to add an identity loss which ensures that the mapping is consistent with the color of the input image. The two mapping functions $G$ and $F$ learn the identity function in case a sample from the real distribution represents the input image. This loss is described by:
\begin{align}
\begin{split}\label{eq:idt}
    L_{identity}\left(G,F,X,Y\right) ={}& \mathbb{E}_{y\sim p_{data}(y)}\left[||G(y)-y||_1\right]+\\
         & \mathbb{E}_{x\sim p_{data}(x)} \left[||F(x) - x||_1\right].
\end{split}
\end{align}
The complete objective function is therefore obtained as: 
\begin{align}
\begin{split}
    L \left( G,F,D_x,D_y,X,Y \right) ={} & L_{GAN} \left( G,D_y,X,Y \right) + \\
                      & L_{GAN} \left( F,D_x,Y,X \right) + \\
                      & \lambda L_{cyc} \left( G,F,X,Y \right) + \\
                      & L_{identity} \left( G,F,X,Y \right)
\end{split}    
\end{align}
\citep{Zhu2017-pb}.
\subsection{Multi-domain stain normalization with MultiStain-CycleGAN}
\begin{figure*}[!t]
\centering
\includegraphics[scale=.15]{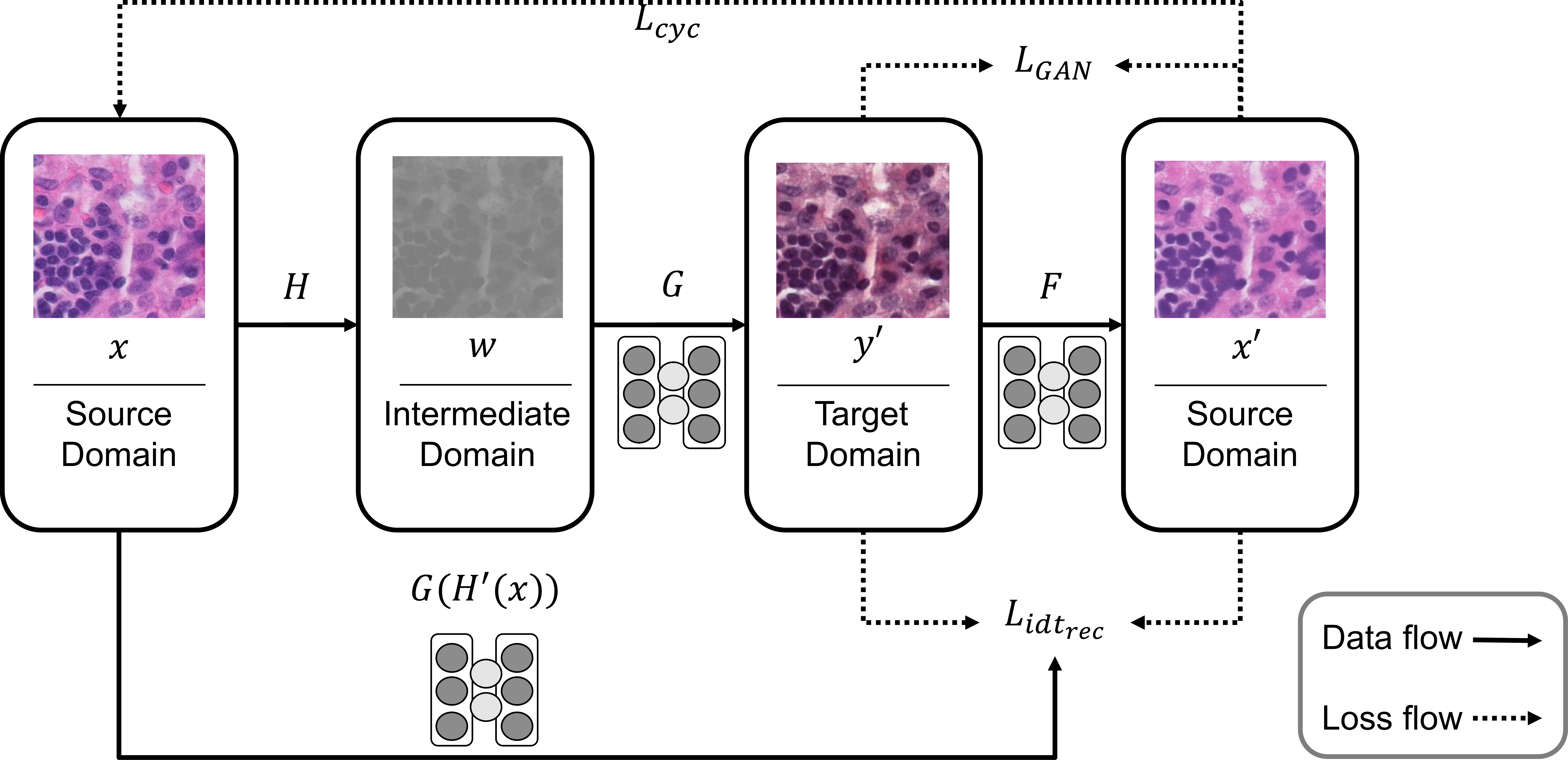}
\caption{
Overview of the MultiStain-CycleGAN. Images $x$ from a source domain  will be mapped to an intermediate domain by a function $H$. $H$ consists of a color augmentation function and a grayscale conversion.   
The generator $G$ then transforms the gray image $w$ into the target domain. This process, including projecting $y'$ into the intermediate domain, is repeated for the normalized image $y'$ again, to reconstruct the original image. The second path has been omitted for clarity. Further, instead of feeding the network a real image from the respective domain for calculating the identity loss used by Zhu et al., a reconstruction task of unaugmented gray images $H'(x)$ is done. The intermediate domain allows to normalize any H\&E stains without having to re-train the model.
}
\label{fig:principle}
\end{figure*}
To convert CycleGAN into a many-to-one approach, we performed several modifications. To reduce the space of possible inputs and thus simplify the problem, the images are converted to 3-channel grayscale images. Thus, the new task for the two mapping functions $G$ and $F$ is the reconstruction of the RGB images of the respective domain from the gray-converted images. We hypothesize that if the reconstruction $F(G(x))$ and $G(F(y))$ is sufficiently good, the loss of information due to gray scale conversion will be compensated by the model and thus be valid. 
In order to increase the variance of the input data and thus improve the generalization ability of the generators, an augmentation function is applied. The augmentation function $H$ then is obtained from the color augmentation and the gray value conversion. This function is essential to later normalize data outside the distribution of the raw training data. Depending on the intensity of the augmentation, more or less information can be lost in the image due to lack of contrast. The task of the generators changes to denoising and recoloring into the target domain. By applying the function $H$, the images are transformed into the intermediate domain $\mathcal{W}$, which represents the input space (see Fig. \ref{fig:principle}). The use of the intermediate domain $\mathcal{W}$ allows us to normalize a large variation of input images at inference time. Since the original identity loss task is no longer relevant in this setup, the additional task of reconstructing unaugmented grayscale images was added instead. This additional task allows to compensate for the noisy images that may result from strong contrast augmentations, thus focusing $G$ and $F$ on the normalization of color instead of the denoising task. This domain faithful reconstruction loss of the gray converted input image $x’$, $y’$ and the original image $x$, $y$ is: 
\begin{align}
\begin{split}\label{eq:idt_rec}
    L_{idt_{rec}}(G,F,X,Y) ={}& \mathbb{E}_{y\sim p_{data}(y)} \left[\|G(y') - y\|_1\right]+\\
         & \mathbb{E}_{x\sim p_{data}(x)} \left[\|F(x') - x\|_1\right] .
\end{split}
\end{align}
Thus, the complete objective function for our model results in:
\begin{align}
    \begin{split}
        L\left(G,F,D_x,D_y,X,Y\right) ={}  &  L_{GAN}\left(G,D_y,X,Y\right)+\\  
                                & L_{GAN}\left(F,D_x,Y,X\right)+\\
                                & \lambda_{cyc} L_{cyc}\left(G,F\right)+\\
                                & \lambda_{idt} L_{idt_{rec}}\left(G,F\right) .
    \end{split}
\end{align}
For our experiments, we choose $\lambda_{cyc}=10$ and $\lambda_{idt}=0.5$ as in the original implementation \citep{Zhu2017-pb}. The two generators $G$ and $F$ are each an adapted U-Net, which has proven to be a very effective architecture in various medical tasks \citep{Ronneberger2015-gz}. The network contains several downsampling, and respective upsampling blocks, which double the number of filters and halve the image dimensions respectively and vice versa. For our evaluation, we implemented a tile size of $256 \times 256$ and a filter count for the innermost block of $32$ proved effective. In the case of a downsampling block, the blocks consist of convolution layers and a leaky ReLu activation function with a slope $a = 0.2$ and an instance normalization layer described by \citep{Ulyanov2016-on}. In the case of an upsampling block, it consists of a transpose convolution layer, a ReLu activation function and an instance normalization layer. 
The two discriminators are PatchGANs proposed by \citep{Isola2017-nl}, their loss can be interpreted as a kind of style loss. In addition, spectral normalization introduced by \citep{Miyato2018-cx} was used as a normalization layer to stabilize the training. The respective discriminator consists of three blocks, each consisting of a convolution layer, a leaky ReLu activation function and a normalization layer. The filter numbers increase quadratically with the depth of the discriminator. The least squares GAN (LSGAN) loss is used as loss, which allows a better image quality for generated images \citep{Mao2017-jg}. To further stabilize the training, and reduce oscillations \citep{Goodfellow2016-fn}, we implemented an image buffer as proposed by \citep{Shrivastava_undated-te}, which includes a history of generated images, of size 50. Since in our experiments we noticed a tendency of the discriminator loss to converge to zero, we introduced an update threshold to avoid this, which prevents a gradient update as soon as one of the discriminator losses falls below a threshold value. The threshold was set to 0.1.\\
For our training, we used the ADAM optimizer \citep{Kingma2014-kl} with a learning rate of $10^{-5}$ with a linear decay over 50 Epochs. We trained the MultiStain-CycleGAN for 100 epochs in total on a Nvidia V100. For the color augmentation we used the following factors: Saturation 0.75, brightness 0.75, contrast 0.5. Since we subject our images to gray conversion, changing the hue value has no effect on the output image. These parameters were determined empirically by visual inspection of test images. They were chosen in such a way that the morphology of the structures is largely preserved. 
\subsection{Image quality and domain shift metrics}
For the evaluation of the image quality and the measurement of the domain shift of our normalization method, we chose the Structural Similarity (SSIM) Index \citep{Wang2004-ic} and the Fréchet Inception Distance (FID) \citep{Heusel2017-bc}, respectively. These metrics are intended to help evaluate the quality of the generated images and quantify the change in domain shift.
\subsubsection{Fréchet inception distance}
\label{subsub:fid}
In order to be able to make a statement about the domain shift before and after normalization and to evaluate different stain normalization methods with respect to their ability to reduce the domain gap, we decided to use the FID described by Heuse et al. The FID is an improvement over the Inception Score proposed by \citep{Salimans2016-qe} in terms of consistency with human perception as the disturbance of the image increases. We chose this metric because of its frequent application in generative tasks and the extensive evaluation of the method \citep{Xu2018-hx,Lucic2017-qn}. The FID represents the difference between two Gaussians consisting of the features of an inception model. The FID is given by:
\begin{equation}
    \text{FID}\left(\mu_X,\mu_Y,C_X,C_Y\right) = \|\mu_X-\mu_Y\|+ Tr\left(C_X+C_Y-2\sqrt{C_XC_Y}\right).
\end{equation}
Where the mean and covariance $\mu_X, C_X$ corresponds to the Gaussian of the generated data and $\mu_Y, C_Y$ corresponds to the Gaussian of the real world data. The FID is zero in case of matching images. \\
Due to the issue described by \citep{Liu2018-kq} that using an ImageNet model to project generated images of domains unrelated to ImageNet into feature space can be ineffective. Following the suggestion of Liu et al., we used the model of \citep{Ciga2022-fd} as a domain-specific encoder, which was trained on several histopathology datasets. The FID correlates with human visual perception and measures of how large the perceived difference is between images from two different distributions. In the stain normalization use case, a high FID means that there is a large domain shift. This domain shift, and thus the FID, should be reduced by stain normalization methods. 
\subsubsection{Structural Similarity Index}
To compare the perceived structure before and after normalization, we utilize the Structural Similarity Index proposed by Wang et al. for our evaluation. The SSIM index compares two images in terms of their similarity and image quality. The SSIM index is 1 in case of two identical images. In contrast to the use of the Mean Squared Error (MSE) for the comparison of images, the SSIM index calculates contrast, structure and luminance separately and then combines them.  We chose this metric because it has already been used in stain normalization scenarios \citep{Hoque2021-bx, Shaban2019-ce} and due to the high importance of keeping structural features unaffected by normalization. Preserving the structure after normalization is essential, otherwise the morphology of the cells is altered. This can lead to errors in classification, as will be shown in chapter \ref{sec:results}. The SSIM is given by:
\begin{equation}
    \text{SSIM}\left(x,y\right) = \frac{\left(2\mu_x\mu_y+C_1\right)\left(2\sigma_{xy}+C_2\right)}{\left(\mu_x^2+\mu_y^2+C_1\right)\left(\sigma_x^2+\sigma_y^2+C_2\right)}.
\end{equation}
Where $\mu_x$ and $\sigma_x^2$ correspond to the mean and variance of the image $x$, respectively. $\sigma_{xy}$ corresponds to the covariance of the images $x$ and $y$, and $C_1$ and $C_2$ are constants to stabilize the denominator.
\subsection{Network-based metrics}
Image-based similarity metrics such as FID and SSIM can give clues about the visual similarity of the normalized pictures. However, they do not evaluate how well downstream classifiers perform on the normalized images. As the goal of stain normalization is to at least maintain task performance while ideally obscuring the origin of the slide, we additionally evaluate our approach with network-based metrics. 
\subsubsection{Domain classifier}
\label{susubsec:tissue_classifier}
To verify that the respective normalization methods are able to disguise the tissue source site, we use an Xception classifier proposed by \citep{Chollet2017-gi} pre-trained on ImageNet to follow \citep{Howard2021-lw}. We measure accuracy because this is a balanced test dataset and we are interested in the clinic an image originates from. As described in chapter \ref{sec:materials}, we applied a 5-fold cross-validation to obtain an approximate distribution of accuracy and exclude outliers. Each model was trained with a learning rate of $10^{-4}$ and a batch size of 32 for 50 epochs. Due to very large performance differences of the discriminator with different levels of color augmentations in training, we decided to use a high intensity of color augmentation. We used color jitter with the parameters brightness: 0.7, contrast: 0.7, saturation: 0.7, hue: 0.5.
\subsubsection{Tumor classifier}
For tumor classification, we utilized a ResNet18 pre-trained on ImageNet. Analogous to the domain classifier, we trained the classifier with a learning rate of $10^{-4}$ and a batch size of 32 for 50 epochs. Again, we chose accuracy as the target metric  due to the testset having balanced classes, as described in chapter \ref{sec:materials}. Also with this classifier we could see very large performance differences depending on the intensity of the color augmentation that was used. Thus, we employed  the same parameters as described for the domain classifier in \ref{susubsec:tissue_classifier}.

\section{Experimental setup}
\label{sec:experimental_setup}
\subsection{Normalization}
\label{subsec:normalization}
\begin{figure}[!t]
\centering
\includegraphics[width=.45\textwidth]{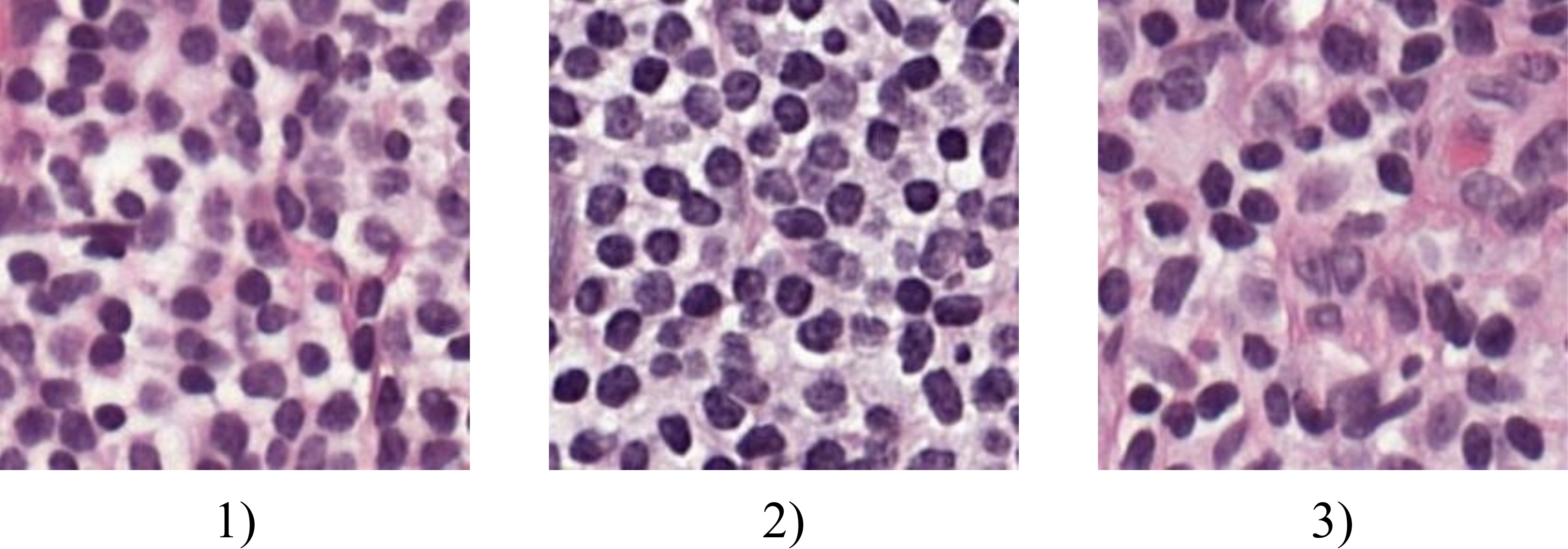}
\caption{The target domain representative template images from three different slides used for the template-based methods.}
\label{fig:templates}
\end{figure}
\begin{table*}[!ht]
\centering
\caption{Estimated mean and standard deviation of domain and tumor classifier accuracy for the different normalization methods analyzed. Furthermore, the mean SSIM index and its standard deviation as well as the FID for the tumor classifier dataset are shown.}
\vspace{5mm}
\label{tab:results_short}
\begin{tabular}{@{}llllll@{}}
                    & \parbox{3cm}{Domain Classifier\\ Accuracy} $\downarrow$ & \parbox{2.5cm}{Tumor Classifier\\ Accuracy} $\uparrow$& \parbox{2cm}{Sustained \\ performance}  & SSIM Index $\uparrow$ & FID $\downarrow$ \\ \toprule
Unnormalized        &$0.952_{\pm0.002}$& $0.901_{\pm0.002}$& - &$1$&$41.389$\\ \midrule
Macenko Template 1  &$0.818_{\pm0,016}$&$0.892_{\pm0.003}$& - &$0.907_{\pm0.089}$&$59.247$\\
Macenko Template 2  &$0.634_{\pm0.029}$&$0.832_{\pm0.008}$& - &$0.753_{\pm0.13}$&$50.550$\\
Macenko Template 3  &$0.829_{\pm0.012}$&$0.887_{\pm0.003}$& - &$0.897_{\pm0.099}$&$62.010$\\ \midrule
Reinhard Template 1 &$0.905_{\pm0.009}$&$0.900_{\pm0.005}$& \checkmark &$0.861_{\pm0.108}$&$35.310$\\
Reinhard Template 2 &$0.910_{\pm0.007}$&$0.904_{\pm0.006}$& \checkmark &$0.818_{\pm0.121}$&$41.720$\\
Reinhard Template 3 &$0.911_{\pm0.006}$&$0.899_{\pm0.005}$& \checkmark &$0.889_{\pm0.099}$&$33.910$\\ \midrule
Vahadane Template 1 &$0.852_{\pm0.019}$&$0.892_{\pm0.003}$& - &$0.898_{\pm0.096}$&$51.150$\\
Vahadane Template 2 &$0.605_{\pm0.040}$&$0.650_{\pm0.018}$& - &$0.674_{\pm0.149}$&$63.590$\\
Vahadane Template 3 &$0.855_{\pm0.012}$&$0.890_{\pm0.003}$& - &$0.900_{\pm0.094}$&$50.060$\\ \midrule
\textbf{MultiStain-CycleGAN}   &$0.701_{\pm0.016}$&$0.900_{\pm0.004}$& \checkmark &$0.957_{\pm0.034}$&$40.874$     \\ \bottomrule
\end{tabular}
\end{table*}
For all experiments, we trained our MultiStain-CycleGAN to learn the transformation $G: \text{CWZ} \to \text{LPON}$ and $F: \text{LPON} \to\text{CWZ}$. For simplicity, we examine only the normalization $F$. Due to the huge memory requirements of WSIs, we were forced to perform the evaluation in a tile-by-tile manner. Thus, in order to analyze all the criteria under consideration, both the domain dataset and the tumor classifier dataset have to be normalized using the normalization methods under investigation. To place our method in the context of existing literature, we compare it with other common normalization methods, which can be used in a multi-domain manner. We chose the methods of \citep{Macenko2009-yu}, \citep{Reinhard2001-qk} and \citep{Vahadane2016-kj} because of their frequent use in stain normalization in histopathology. Due to the template-based nature of these methods, we selected three representative templates shown in Fig. \ref{fig:templates} for these. This leads to three normalized datasets for each of the template-based approaches for the respective task. For the Machenko normalization and the Reinhard and Vahadane method, we utilized the implementations from torchstain\footnote{https://github.com/EIDOSLAB/torchstain} and staintools\footnote{https://github.com/Peter554/StainTools} respectively.  
\subsection{Deep learning classifier}
\label{subsec:classifier_eval}
We trained each of the classifiers five times using different seeds. We trained our models with four different augmentation intensities to analyze the behavior of the studied methods considering color augmentation.  Since we use tiles of size $256 \times 256$ as input for the normalization methods as described in chapter \ref{sec:materials}, they were brought to the size $224 \times 224$ using a center crop in order to use the pretrained ResNet18 and the Xception model. For each task, the accuracy for each of the normalized and unnormalized datasets is determined for each of the five models.
\subsection{FID and SSIM index}
\label{subsec:fid_ssim}
We likewise use an open-source implementation\footnote{https://github.com/mseitzer/pytorch-fid} to calculate the FID and have modified it as described in section \ref{subsub:fid} to use a model trained on histopathological data. We calculated the FID for both the normalized domain and tumor classifier datasets by computing the distance between 40 000 random tiles from the 10 WSIs with lesion-level annotations of the target domain and the respective normalized dataset.\\
The SSIM index is calculated using image pairs compared to the FID. Here, for each of the normalized datasets, the SSIM index is calculated using the normalized image and the non-normalized image. Thus, for each of the datasets, we estimate a mean value and the associated standard deviation. For the calculation we use the implementation of scikit-image \citep{Van_der_Walt2014-br}.
\section{Results}
\label{sec:results}
\begin{figure*}[!ht]
 \centering
     \includegraphics[width=\textwidth]{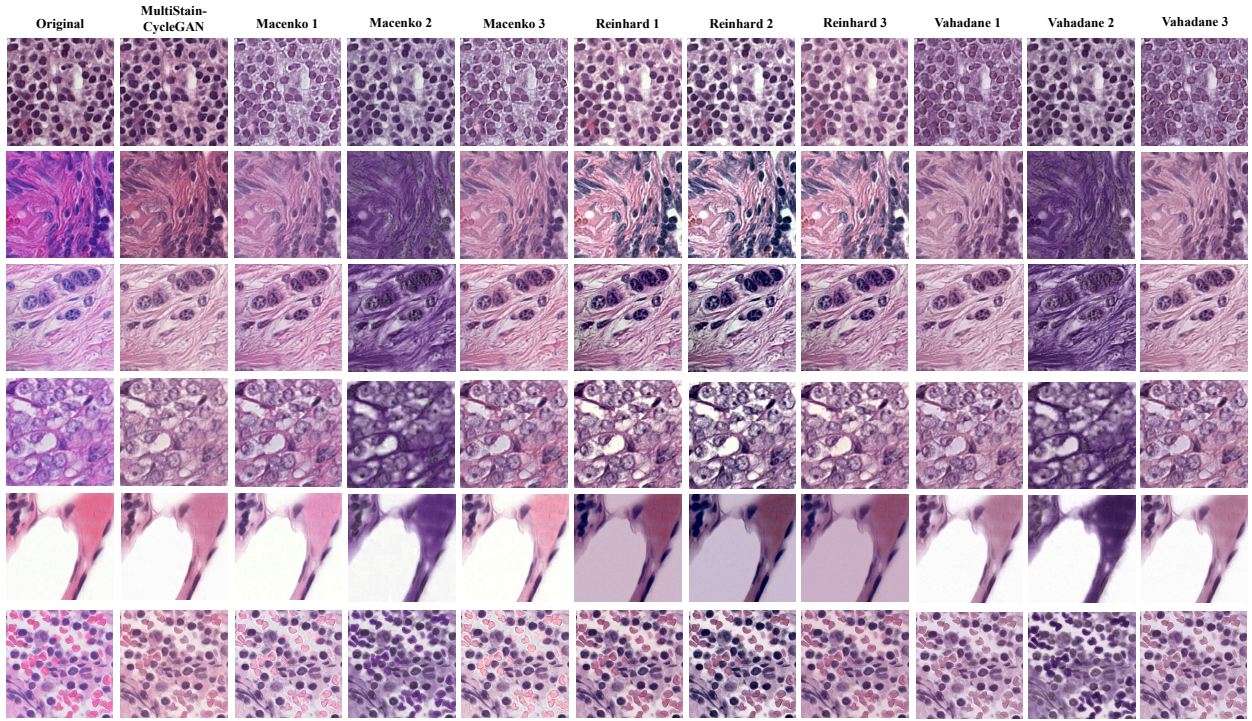}
     \caption{Examples of normalization of the different methods studied. Tiles are taken from slides of all centers. Our method achieves very good stain adaptation. Furthermore, the problems of the template-based normalization are shown by very different results depending on the template. }
     \label{fig:normed_images}
\end{figure*}
\begin{figure*}[!ht]
 \centering
     \includegraphics[width=0.9\textwidth]{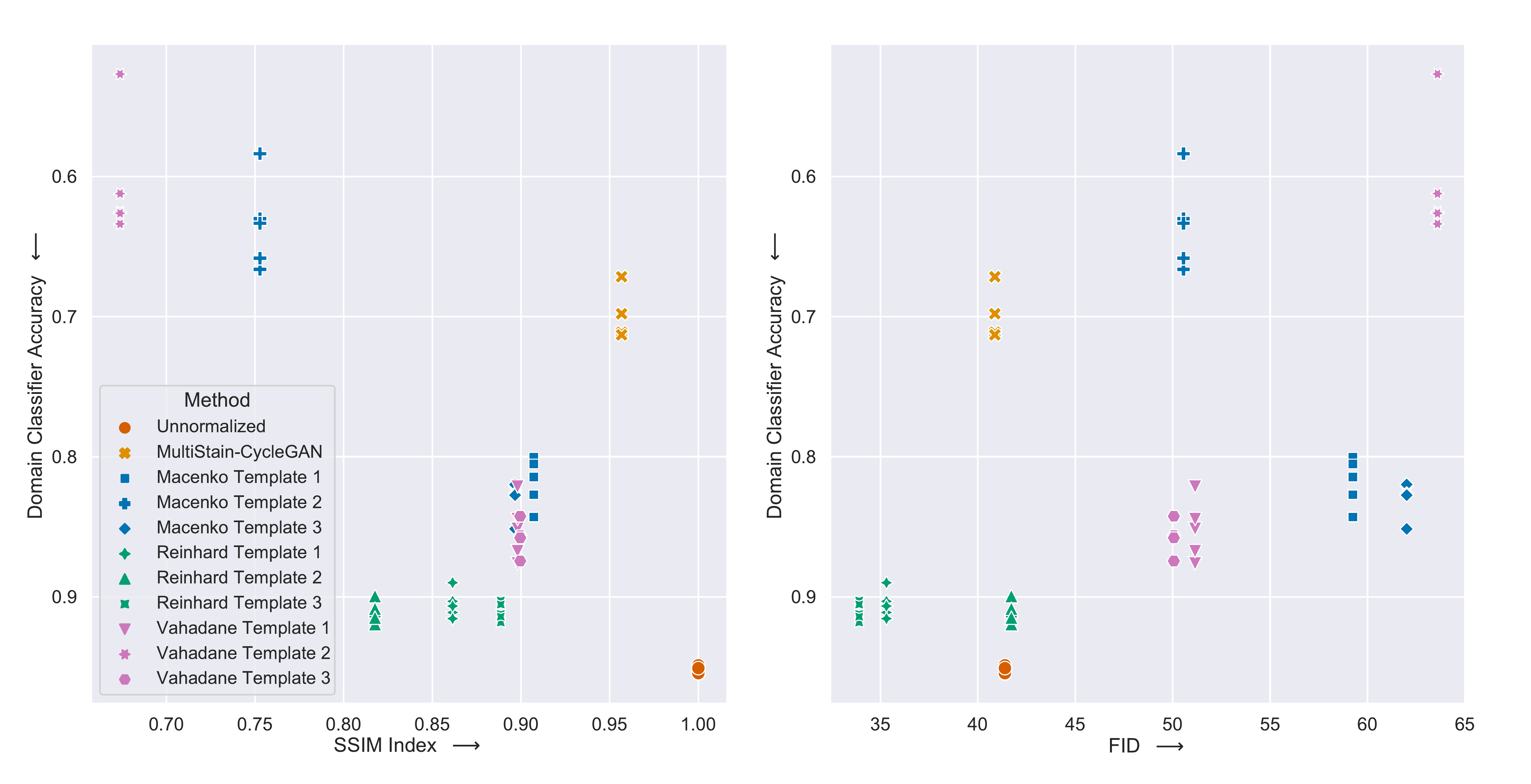}
     \caption{The behavior of the domain classifier under different normalization methods over the two metrics SSIM index and FID. For each method, the five trained models are visualized. MultiStain-CycleGAN achieves the highest SSIM index while performing well in fooling the domain classifier. Macenko 2 and Vahadane 2 acquire the low accuracy of the domain classifier with a strong perturbation of the image content.}
     \label{fig:tissue_site_acc}
\end{figure*}
\begin{figure*}[!ht]
 \centering
     \includegraphics[width=\textwidth]{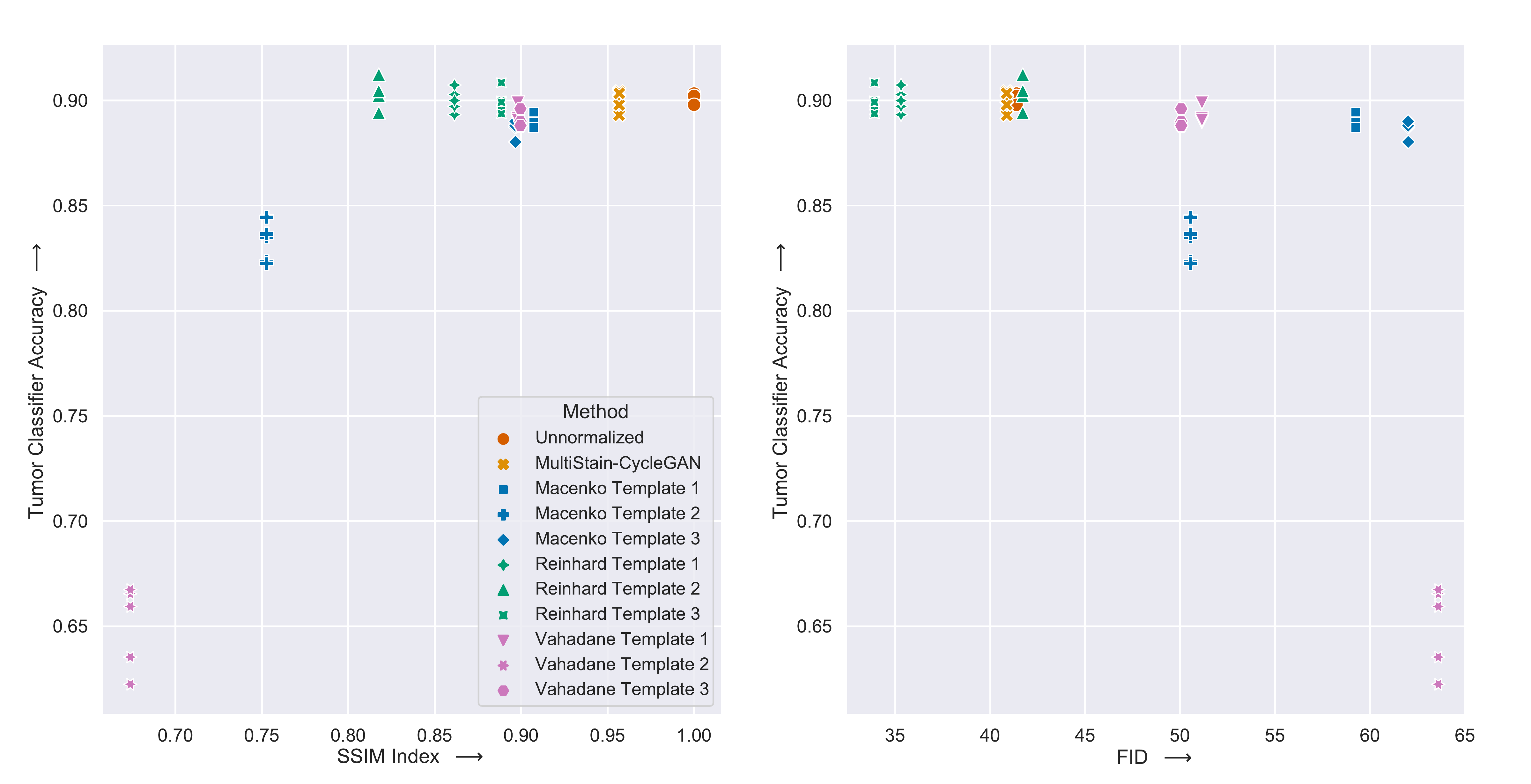}
     \caption{The dependence of tumor classifier accuracy on SSIM index on the left and FID on the right. Each method is represented by the five models trained. It is shown that with decreasing SSIM index, tumor classifier performance decreases significantly. This is possibly due to loss of contrast and the associated structural changes and loss of information in the images. The result is similar for the FID, where accuracy decreases with increasing FID.}
     \label{fig:tumor_acc}
\end{figure*}
A summary of the results is given in table \ref{tab:results_short}. This includes the classifier results averaged over each of the five models for the heavy augmentation level, the mean SSIM index, and the FID. In addition, we introduce a check that indicates whether the ranges $\mu \pm \sigma$ of tumor classifier accuracy of the respective methods overlap with or exceed the range of the unnormalized case. Where $\mu$ denotes the estimated mean and $\sigma$ denotes the estimated standard deviation of the tumor classifier accuracy. This is then referred to as sustained performance. An overview of the normalization results of the different methods studied is provided by Fig. \ref{fig:normed_images}. This shows the original images and the normalized variant in each case. The images are from all of the centers available in the Camelyon17 dataset. For comparison, images of the target domain CWZ are given in \ref{fig:examples_images}. More tiles normalized with our method can be seen in \ref{fig:normalization_examples}. Here, the tiles are from the untrained centers RST, UMCU and RUMC. In the following, the results of domain classification will be discussed in more detail.    
\subsection{Domain classification}
\label{subsec:tissue_site_results}
We examined the results of the tissue site classifier, or domain classifier, for the different normalization methods, as well as for the unnormalized case. A detailed overview of the different classification results of the domain classifier are given in table \ref{tab:detailed_domain_results}. \\ 
The domain could be estimated with very high accuracy of 0.952 in the unnormalized case. The Reinhard normalization still has a high accuracy of over 0.9 in all cases. The lowest accuracy was achieved by the Vahadane and Macenko normalization for template 2 with 0.605 and 0.634, respectively, followed by MultiStain-CycleGAN with an accuracy of 0.701. Fig. \ref{fig:tissue_site_acc} shows the accuracy of the domain classifier over the two metrics SSIM index and the FID. In the left plot, as well as in table \ref{tab:results_short}, it can be seen that the two methods that result in the lowest accuracy of the domain classifier also have a very low SSIM index. Further, our presented method achieves the highest SSIM index, with a value of 0.957. \\ 
In the right plot in Fig. \ref{fig:tissue_site_acc}, the accuracy of the domain classifier over the FID is shown. Again, the outlier character of the two methods Macenko Template 2 and Vahadane Template 2 can be seen. The two methods have by far the highest FID. Our method is placed in the middle and only slightly below the unnormalized data. In general, a trend emerges here, which shows that with increasing FID, the accuracy of the domain classifier also decreases, although our approach is somewhat out of line here.
\subsection{Tumor classification}
\label{subsec:tumor_acc_results}
The evaluation of the tumor classifier was performed analogously to the domain classifier. Again, a detailed listing of the results is given in the appendix in table \ref{tab:detailed_tumor}. The accuracy of tumor classification is relatively high at about 0.9 in all studied normalization methods as well as in the unnormalized case, with the exception of the methods Macenko Template 2 and Vahadane Template 2. These two methods have a significant loss in accuracy. Thus, for this task, no improvement in classifier performance is shown by applying the studied stain normalizations. \\
Fig. \ref{fig:tumor_acc} on the left shows tumor classifier accuracy as a function of the SSIM index. It can be seen that most methods have a very similar performance compared to the unnormalized case. Furthermore, several methods decrease the accuracy moderately to strongly compared to the unnormalized case.  This case can be observed especially when the SSIM index decreases strongly. On the right side of Fig. \ref{fig:tumor_acc}, tumor classifier accuracy over FID is visualized. The tumor classifier accuracy decreases with increasing FID. However, the trend here is less clear than in the case of the left plot.

\section{Discussion}
\label{sec:discussion}
\begin{figure}[ht]
 \centering
     \includegraphics[width=0.45\textwidth]{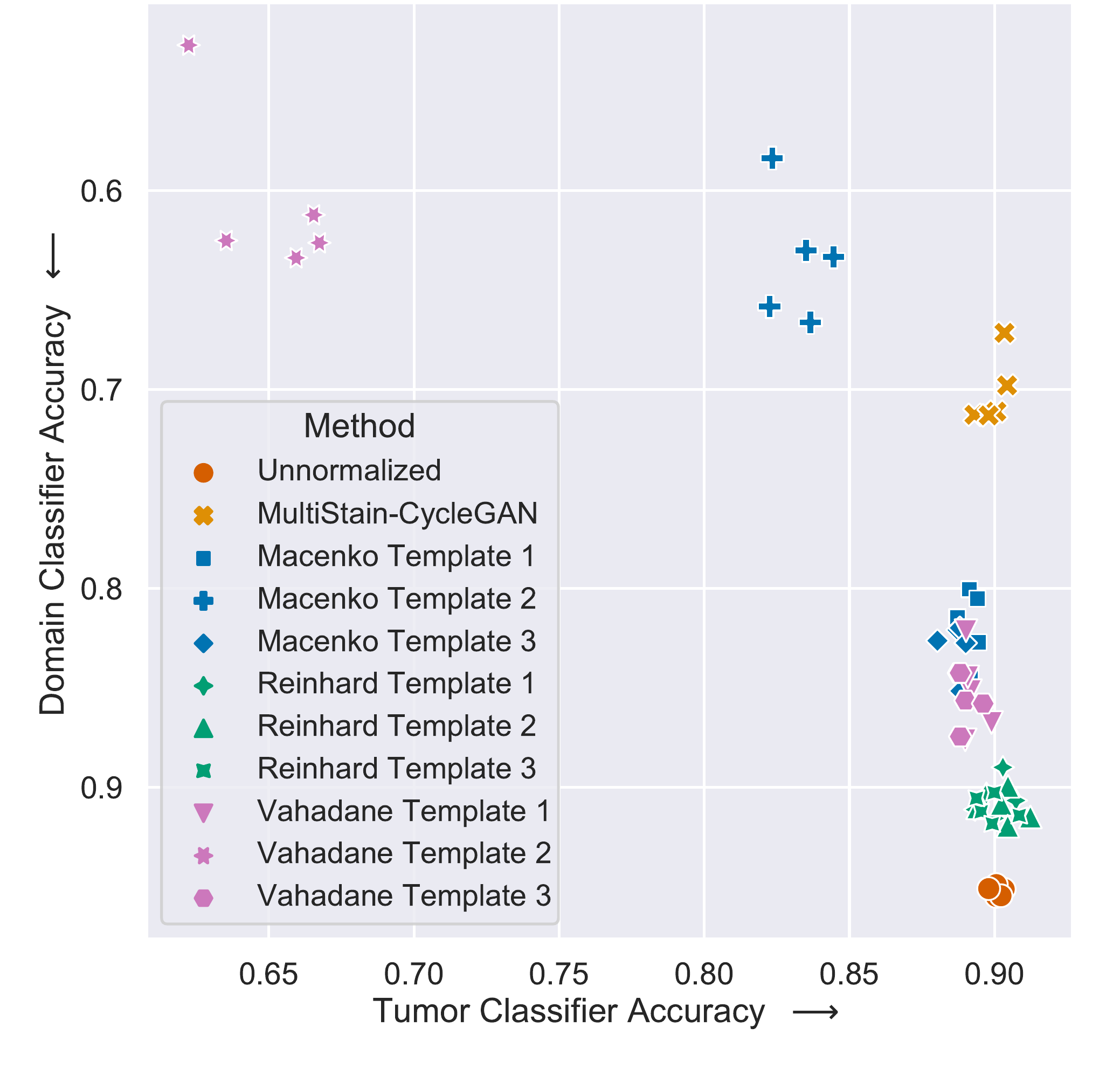}
     \caption{The accuracy of the domain classifier over the tumor classifier accuracy of the different methods. MultiStain-CycleGAN is the only method that is able to both fool the domain classifier and maintain tumor classifier accuracy. Despite its frequent use, Macenko normalization in our case decreases tumor classifier accuracy in all cases. The Reinhard normalization provides a normalization that is very consistent, but does not offer the capability to fool the domain classifier. }
     \label{fig:domain_over_tumor_acc}
\end{figure}

As shown in Fig. \ref{fig:tumor_acc}, saturation effects occur for both metrics. It can be seen that the performance of the tumor classification is only significantly reduced when the SSIM index falls below a threshold. Similar results can be observed for FID, but the trend is somewhat more continuous. \\
As expected, a high FID and low SSIM index results in a significant performance loss of the tumor classifier. This possibly results from the fact that the two methods reduce the contrast too much, altering the morphology of the tissue. It follows that this is a severe loss of information. Due to the loss of essential information, the domain can no longer be estimated by the domain classifier. Interestingly, despite a mid-range FID, our method is significantly more capable than other methods in fooling the domain classifier, except for the two methods Macenko Template 2 and Vahadane Template 2. Thus, our method can beat the other methods by the highest SSIM index and comparatively very good ability to fool the domain classifier. This behavior is again supported by Fig. \ref{fig:domain_over_tumor_acc}, which shows a clear gap in domain classifier accuracy between our method and the other normalization methods, and at the same time a clear gap in tumor classifier accuracy to the two outlier methods. \\
Despite only slightly lower FID compared to the unnormalized case, our approach significantly lowered the accuracy of the domain classifier. Here, the visual perception and the FID drift apart, as one can see clear differences before and after normalization in Fig. \ref{fig:normed_images}. Thus, in our case, we see the FID only conditionally suitable to quantify domain shifts for histopathological data.
Furthermore, the metrics show that our method is able to normalize stainings that are not contained within the training set and thus can normalize different domains with one model without the need for retraining. This is supported by the high tumor classifier accuracy despite distribution shifts through the different centers showing how good the ability of our method is to normalize unseen domains in training.  
We could not observe any significant improvement of the downstream task, in our case tumor classification, in the case of heavy data augmentation. This is partly consistent with the findings of \citep{Tellez2019-fo}. Here, however, it could be due to the task being too trivial to solve. Based on the tables \ref{tab:detailed_domain_results} and \ref{tab:detailed_tumor}, it can be seen that it is essential to investigate stain normalization methods considering color augmentation. The results show that when using no to moderate color augmentation, our normalization method can achieve high improvements, which however become smaller when using strong color augmentation. Thus, the domain classifier can be fooled much less with heavy color augmentation applied in training and the accuracy increases by more than 30\% compared to the use of no color augmentation. 
A similar picture emerges for tumor classification, where our normalization provides an accuracy gain of over 10\% in the case of no color augmentation. In the case of medium color augmentation, we still achieve a slightly increased performance, which then disappears in the case of heavy color augmentation.\\
Another point that the results show is that template-based approaches, depending on the template chosen, can deliver very different solutions and associated performances with very high variances. Thus, the choice of an inappropriate template can lead to performance losses. GAN-based approaches have an advantage here, since they take the entire training dataset into account and can thus lead to more consistent solutions.\\
In the end, we can say that we can partially reproduce the results of \citep{Howard2021-lw} and extend them with a GAN-based approach. Also, our results show that despite the use of commonly used stain normalization methods, the domain can be predicted with high accuracy.  Our method is the exception here, as it results in a significant loss of domain classifier accuracy while maintaining high image quality. 

\section{Conclusion}
\label{sec:conclusion}
We have presented MultiStain-CycleGAN, a new approach to stain normalization based on a modified CycleGAN, using an intermediate domain which works for multiple unseen stainings without the need to retrain and is thus multi-domain capable. We have extensively compared our approach with several commonly used normalization methods. We have intensively analyzed the different methods using different metrics and augmentation levels to understand the behavior of the methods under investigation. It has been shown that our method does not suffer from the problems of template-based approaches, and at the same time, in contrast to conventional GAN-based approaches, training a single model is sufficient. Furthermore, our method is best able to fool the domain classifier while providing the best image quality and high performance in the downstream task. In doing so, this work is a step towards disguising the origin of the tissue sections and reducing bias through the different domains. Continuing this work, other stainings such as IHC can be investigated. Furthermore, more complex downstream tasks should be analyzed, since in the case of our task none of the normalization methods led to an improvement in performance.

\section*{Acknowledgments}
The research is funded by the \textit{Ministerium für Soziales und Integration}, Baden Württemberg, Germany.

\bibliographystyle{unsrtnat}
\bibliography{refs}  %%% Uncomment this line and comment out the ``thebibliography'' section below to use the external .bib file (using bibtex) .




% 	\newblock {\em arXiv preprint arXiv:1804.09028}                 , 2018.

% \end{thebibliography}
\appendix
\section{Example images}
\label{fig:examples_images}
\begin{figure*}[tbh]
    \centering
    \includegraphics[width=.75\textwidth]{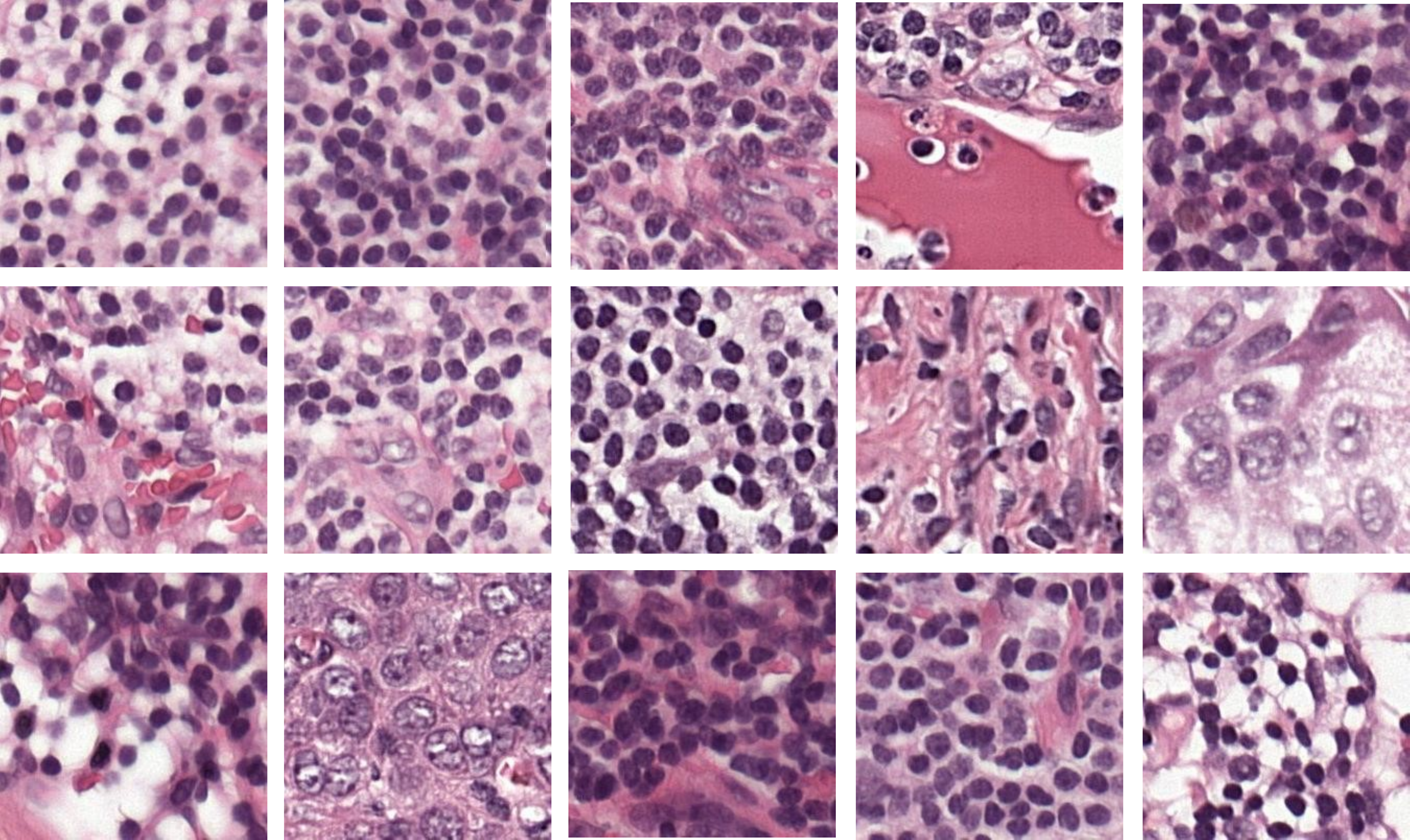}
    \caption{Representative sample images from CWZ.}
    \label{fig:center0_examples}
\end{figure*}
\begin{figure*}[tbh]
    \centering
    \includegraphics[width=.75\textwidth]{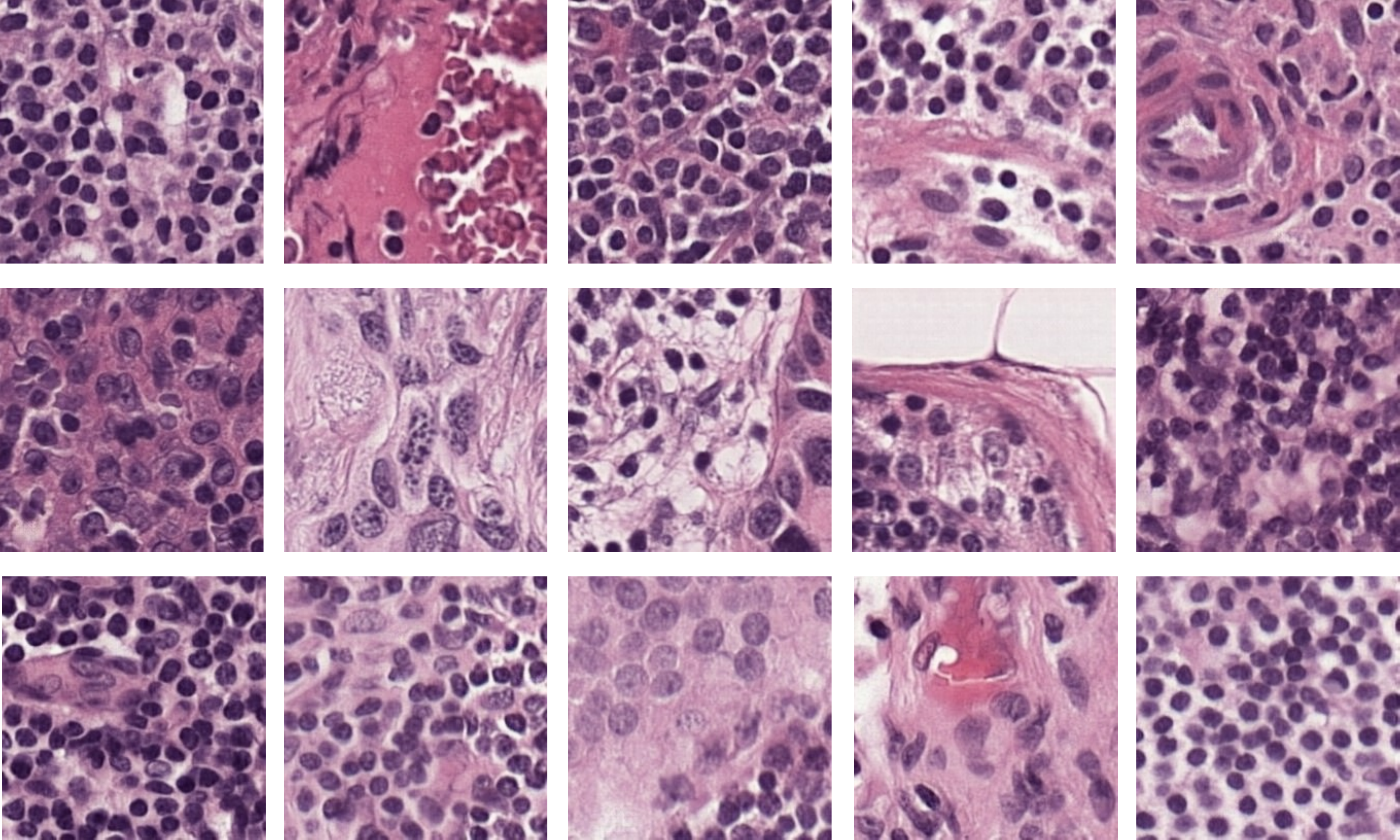}
    \caption{Example images of our proposed normalization. Each row shows images of one of the three untrained centers RST, UMCU and RUMC.}
    \label{fig:normalization_examples}
\end{figure*}
\newpage
\section{Detailed results}
\label{apx:detailed_results}
\begin{table*}[tbh]
\centering
\caption{Detailed results of the different domain classifications for the used augmentation levels and the different models.}
\label{tab:detailed_domain_results}
{\small %
\begin{tabular}{@{}lllllll@{}}
\toprule
Augmentation & Normalization       & 1     & 2     & 3     & 4     & 5     \\ \midrule
             & Unnormalized        & 0.978 & 0.980 & 0.979 & 0.976 & 0.978 \\
             & \textbf{MultiStain-CycleGAN}            & 0.332 & 0.310 & 0.318 & 0.327 & 0.314 \\
             & Macenko Template 1  & 0.524 & 0.482 & 0.575 & 0.553 & 0.561 \\
             & Macenko Template 2  & 0.304 & 0.217 & 0.279 & 0.394 & 0.351 \\
             & Macenko Template 3  & 0.520 & 0.493 & 0.561 & 0.579 & 0.498 \\
None         & Reinhard Template 1 & 0.338 & 0.256 & 0.325 & 0.421 & 0.292 \\
(0,0,0,0)              & Reinhard Template 2 & 0.377 & 0.267 & 0.318 & 0.481 & 0.303 \\
             & Reinhard Template 3 & 0.341 & 0.262 & 0.332 & 0.393 & 0.302 \\
             & Vahadane Template 1 & 0.383 & 0.347 & 0.390 & 0.428 & 0.386 \\
             & Vahadane Template 2 & 0.218 & 0.209 & 0.212 & 0.255 & 0.268 \\
             & Vahadane Template 3 & 0.435 & 0.380 & 0.445 & 0.488 & 0.404 \\ \midrule
             & Unnormalized        & 0.958 & 0.965 & 0.959 & 0.962 & 0.958 \\
             & \textbf{MultiStain-CycleGAN}            & 0.384 & 0.445 & 0.375 & 0.341 & 0.331 \\
             & Macenko Template 1  & 0.746 & 0.694 & 0.733 & 0.768 & 0.717 \\
             & Macenko Template 2  & 0.512 & 0.455 & 0.395 & 0.441 & 0.438 \\
             & Macenko Template 3  & 0.769 & 0.736 & 0.773 & 0.788 & 0.757 \\
Light        & Reinhard Template 1 & 0.800 & 0.745 & 0.773 & 0.798 & 0.830 \\
(0.2,0.2,0.2,0.2)             & Reinhard Template 2 & 0.785 & 0.724 & 0.743 & 0.773 & 0.805 \\
             & Reinhard Template 3 & 0.802 & 0.757 & 0.777 & 0.790 & 0.819 \\
             & Vahadane Template 1 & 0.670 & 0.657 & 0.650 & 0.650 & 0.635 \\
             & Vahadane Template 2 & 0.470 & 0.420 & 0.336 & 0.407 & 0.437 \\
             & Vahadane Template 3 & 0.705 & 0.686 & 0.691 & 0.696 & 0.670 \\ \midrule
             & Unnormalized        & 0.954 & 0.959 & 0.951 & 0.962 & 0.946 \\
             & \textbf{MultiStain-CycleGAN}            & 0.570 & 0.568 & 0.647 & 0.590 & 0.492 \\
             & Macenko Template 1  & 0.790 & 0.809 & 0.819 & 0.815 & 0.842 \\
             & Macenko Template 2  & 0.533 & 0.510 & 0.577 & 0.485 & 0.535 \\
             & Macenko Template 3  & 0.833 & 0.837 & 0.828 & 0.844 & 0.850 \\
Medium        & Reinhard Template 1 & 0.863 & 0.870 & 0.833 & 0.871 & 0.845 \\
(0.4, 0.4, 0.4, 0.4)             & Reinhard Template 2 & 0.858 & 0.865 & 0.837 & 0.879 & 0.850 \\
             & Reinhard Template 3 & 0.877 & 0.887 & 0.858 & 0.888 & 0.858 \\
             & Vahadane Template 1 & 0.818 & 0.828 & 0.828 & 0.841 & 0.826 \\
             & Vahadane Template 2 & 0.446 & 0.471 & 0.556 & 0.419 & 0.463 \\
             & Vahadane Template 3 & 0.821 & 0.837 & 0.833 & 0.844 & 0.815 \\ \midrule
             & Unnormalized        & 0.951 & 0.955 & 0.949 & 0.954 & 0.951 \\
             & \textbf{MultiStain-CycleGAN}            & 0.698 & 0.713 & 0.711 & 0.713 & 0.672 \\
             & Macenko Template 1  & 0.843 & 0.800 & 0.805 & 0.814 & 0.827 \\
             & Macenko Template 2  & 0.584 & 0.630 & 0.666 & 0.633 & 0.658 \\
             & Macenko Template 3  & 0.851 & 0.821 & 0.820 & 0.826 & 0.827 \\
Heavy        & Reinhard Template 1 & 0.890 & 0.911 & 0.903 & 0.916 & 0.907 \\
(0.7, 0.7, 0.7, 0.5)             & Reinhard Template 2 & 0.899 & 0.910 & 0.908 & 0.919 & 0.915 \\
             & Reinhard Template 3 & 0.903 & 0.912 & 0.905 & 0.918 & 0.914 \\
             & Vahadane Template 1 & 0.876 & 0.851 & 0.821 & 0.844 & 0.867 \\
             & Vahadane Template 2 & 0.527 & 0.612 & 0.625 & 0.626 & 0.634 \\
             & Vahadane Template 3 & 0.874 & 0.856 & 0.842 & 0.842 & 0.858 \\ \bottomrule
\end{tabular}
}%
\end{table*}
\begin{table*}[]
\centering
\caption{Detailed results of the tumor classification for the used augmentation levels and the different models.}
\label{tab:detailed_tumor}
{\small %
\begin{tabular}{@{}lllllll@{}}
\toprule
Augmentation & Normalization       & 1     & 2     & 3     & 4     & 5     \\ \midrule
             & Unnormalized        & 0.757 & 0.641 & 0.771 & 0.793 & 0.778 \\
             & \textbf{MultiStain-CycleGAN}            & 0.888 & 0.865 & 0.863 & 0.870 & 0.873 \\
             & Macenko Template 1  & 0.820 & 0.700 & 0.797 & 0.858 & 0.825 \\
             & Macenko Template 2  & 0.478 & 0.565 & 0.495 & 0.475 & 0.550 \\
             & Macenko Template 3  & 0.774 & 0.759 & 0.761 & 0.809 & 0.769 \\
None          & Reinhard Template 1 & 0.800 & 0.759 & 0.706 & 0.815 & 0.797 \\
(0,0,0,0)             & Reinhard Template 2 & 0.846 & 0.777 & 0.569 & 0.833 & 0.848 \\
             & Reinhard Template 3 & 0.833 & 0.801 & 0.783 & 0.836 & 0.817 \\
             & Vahadane Template 1 & 0.863 & 0.834 & 0.832 & 0.873 & 0.857 \\
             & Vahadane Template 2 & 0.499 & 0.523 & 0.502 & 0.504 & 0.613 \\
             & Vahadane Template 3 & 0.871 & 0.826 & 0.835 & 0.876 & 0.857 \\ \midrule
             & Unnormalized        & 0.805 & 0.897 & 0.847 & 0.815 & 0.759 \\
             & \textbf{MultiStain-CycleGAN}            & 0.895 & 0.907 & 0.881 & 0.886 & 0.889 \\
             & Macenko Template 1  & 0.861 & 0.900 & 0.878 & 0.886 & 0.857 \\
             & Macenko Template 2  & 0.841 & 0.828 & 0.761 & 0.825 & 0.763 \\
             & Macenko Template 3  & 0.847 & 0.892 & 0.865 & 0.877 & 0.855 \\
Light         & Reinhard Template 1 & 0.884 & 0.906 & 0.896 & 0.893 & 0.877 \\
(0.2,0.2,0.2,0.2)             & Reinhard Template 2 & 0.878 & 0.904 & 0.907 & 0.905 & 0.866 \\
             & Reinhard Template 3 & 0.908 & 0.919 & 0.909 & 0.908 & 0.899 \\
             & Vahadane Template 1 & 0.882 & 0.910 & 0.886 & 0.891 & 0.875 \\
             & Vahadane Template 2 & 0.592 & 0.630 & 0.557 & 0.694 & 0.594 \\
             & Vahadane Template 3 & 0.873 & 0.907 & 0.878 & 0.886 & 0.873 \\ \midrule
             & Unnormalized        & 0.905 & 0.911 & 0.876 & 0.901 & 0.904 \\
             & \textbf{MultiStain-CycleGAN}            & 0.912 & 0.913 & 0.898 & 0.907 & 0.910 \\
             & Macenko Template 1  & 0.909 & 0.890 & 0.881 & 0.888 & 0.896 \\
             & Macenko Template 2  & 0.815 & 0.878 & 0.814 & 0.826 & 0.752 \\
             & Macenko Template 3  & 0.903 & 0.875 & 0.878 & 0.881 & 0.889 \\
Medium        & Reinhard Template 1 & 0.904 & 0.897 & 0.907 & 0.914 & 0.903 \\
(0.4, 0.4, 0.4, 0.4)             & Reinhard Template 2 & 0.902 & 0.903 & 0.918 & 0.920 & 0.897 \\
             & Reinhard Template 3 & 0.912 & 0.902 & 0.907 & 0.917 & 0.908 \\
             & Vahadane Template 1 & 0.907 & 0.896 & 0.887 & 0.900 & 0.904 \\
             & Vahadane Template 2 & 0.622 & 0.666 & 0.600 & 0.651 & 0.629 \\
             & Vahadane Template 3 & 0.904 & 0.891 & 0.881 & 0.896 & 0.902 \\ \midrule
             & Unnormalized        & 0.903 & 0.901 & 0.901 & 0.902 & 0.898 \\
             & \textbf{MultiStain-CycleGAN}            & 0.904 & 0.893 & 0.901 & 0.898 & 0.903 \\
             & Macenko Template 1  & 0.892 & 0.891 & 0.894 & 0.887 & 0.894 \\
             & Macenko Template 2  & 0.823 & 0.835 & 0.837 & 0.845 & 0.823 \\
             & Macenko Template 3  & 0.888 & 0.887 & 0.888 & 0.880 & 0.890 \\
Heavy         & Reinhard Template 1 & 0.903 & 0.893 & 0.897 & 0.900 & 0.907 \\
(0.7, 0.7, 0.7, 0.5)             & Reinhard Template 2 & 0.905 & 0.894 & 0.902 & 0.905 & 0.912 \\
             & Reinhard Template 3 & 0.900 & 0.895 & 0.894 & 0.899 & 0.908 \\
             & Vahadane Template 1 & 0.890 & 0.892 & 0.890 & 0.891 & 0.899 \\
             & Vahadane Template 2 & 0.622 & 0.665 & 0.635 & 0.667 & 0.659 \\
             & Vahadane Template 3 & 0.888 & 0.890 & 0.888 & 0.888 & 0.896 \\ \bottomrule
\end{tabular}
}
\end{table*}

%\section*{Acknowledgments}
%The research is funded by the \textit{Ministerium für Soziales und Integration}, Baden %Württemberg, Germany.

\end{document}